\documentclass[twocolumn,showpacs,preprintnumbers,amsmath,amssymb,aps]{revtex4}
\usepackage{graphicx}

\begin{document}

\title{Three-photon coherence in a ladder-type atomic system}

\author{Heung-Ryoul Noh$^1$\footnote{hrnoh@chonnam.ac.kr} and Han Seb Moon$^2$\footnote{hsmoon@pusan.ac.kr}}

\affiliation{$^1$Department of Physics, Chonnam National University,
Gwangju 500-757, Korea\\ $^2$Department of Physics, Pusan National University, Busan 609-735, Korea}

\begin{abstract}

We present a theoretical study of three-photon electromagnetically induced absorption for a ladder-type three-level atomic system. A probe beam was tuned to the lower line and two counter-propagating, linearly polarized coupling beams were tuned to the upper line. The system can be modeled with a three- (or five-) level scheme when the polarization directions of the coupling beams are parallel (or perpendicular). By calculating the absorption coefficients analytically for the two schemes, we found that the corresponding absorption coefficients were identical except for different transition strengths, and that the primitive scheme embedded in those schemes was a simple four-level scheme.


\end{abstract}

\date{\today}

\maketitle

\section{Introduction}

\label{Introduction}

The atomic coherence generated between dipole-forbidden states due to two laser beams plays an important role in the abrupt enhancement of transmission or absorption. The effects resulting from this coherence are called electromagnetically induced transparency (EIT) \cite{Harris97,EIT} and absorption (EIA) \cite{Akulshin98,Lezama99a,Taichenachev}, and have drawn much interest for their potential application to atomic and laser spectroscopies. Particular applications include light storage \cite{Lukin}, high-resolution spectroscopy \cite{Krishna05}, nonlinear optics \cite{Braje04}, and quantum information \cite{Polzik08}. As in traditional EIA, enhanced absorption is observed when a third laser beam is introduced in two-beam spectroscopy setup, which displays EIT. In this spectroscopic setup, three-photon atomic coherence results in enhanced absorption, rather than the enhanced transmission that aries from two-photon coherence. This enhanced absorption is known to result from constructive interference \cite{Mulchan00,Gong11}. It should be noted that destructive interference with three laser beams can be observed for different transition geometries \cite{Gong11,Yan01}.

Many studies on three-photon spectroscopy are reported in the literature. A narrow resonance was observed in a four-level atomic system \cite{Bason08}. Chanu \textit{et al.} reported the conversion from EIT to EIA by forming an N-type scheme \cite{Chanu01}, which was subsequently explained theoretically by analyzing the existing subsystems \cite{Pandey13}. Ben-Aroya also reported large-contrast absorption resonances in an N-type scheme \cite{Ben11}. An optical clock based on three-photon resonance was also reported \cite{Hong05}. Enhanced absorption with a standing-wave coupling field for a $\Lambda$-type atomic system was observed \cite{Babin03,Bae10}. Zhu \textit{et al.} examined bichromatic EIT and four-wave mixing in $\Lambda$-type atomic systems \cite{Wang03,Yang10}.

Recently, Moon \textit{et al.} reported three-photon electromagnetically induced absorption (TPEIA) in a ladder-type atomic system where a probe beam was tuned to the lower transition line and two counter-propagating coupling beams were tuned to the upper transition line \cite{Jeong}. Calculations within the simple four-level model showed that this absorption phenomenon was due to three-photon coherence. They also studied the relationship between two- and three-photon coherence \cite{YSLee}. As in previous studies \cite{Jeong,YSLee}, where the two coupling beams were linearly polarized in perpendicular directions, three-photon coherence was the main cause of the enhanced absorption. However, when the two coupling beams were polarized in the same direction, it was not obvious whether or not three-photon coherence was established. Enhanced absorption may also have been observed in this case. It should be also noted that TPEIA is different from the conventional EIA where the absorption results from the transfer of the Zeeman coherence of the excited state \cite{Akulshin98,Lezama99a,Taichenachev}. The main cause of the absorption in TPEIA is the three-photon coherence generated between the ground and intermediate states.

We here present a theoretical and analytical study of TPEIA for two schemes, where the coupling-beam polarizations are either parallel or perpendicular to one another. To study TPEIA analytically, these two schemes were modeled using three-level and five-level schemes. After deriving the analytical form of the absorption coefficients in the limit of weak probe intensity, we studied the similarities and differences between the two schemes. These are the new findings as compared to the previous reports \cite{Jeong,YSLee} as well as accurate analytical solutions of the absorption coefficients. This paper is structured as follows. Section \ref{motivation} briefly describes the motivation and experimental findings of our work. The theoretical method of the calculation is presented in Sec. \ref{theory}. Section \ref{results} discusses the absorption spectra averaged over the Doppler-broadened atomic velocity distribution. Finally, we summarize the results.

\section{Motivation}
\label{motivation}

\begin{figure}[thb]
\centerline{\includegraphics[width=8cm]{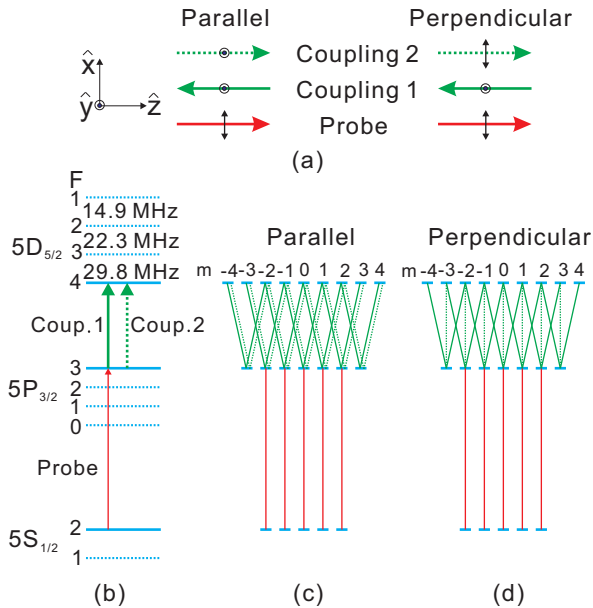}} \caption{(Color online) (a) Simplified schematic diagram for the TPEIA experiment. (b) Energy-level diagram for the $5S_{1/2}-5P_{3/2}-5D_{5/2}$ transition in $^{87}$Rb atoms. The two-coupling beams are linearly polarized in (c) parallel or (d) perpendicular direction.}
\end{figure}

The simplified schematic diagram and the energy-level diagram for the experiment considered are shown in Fig. 1. The probe beam is tuned to the $5S_{1/2}(F=2) - 5P_{3/2}(F'=3)$ transition, and two counter-propagating coupling beams are tuned to the $5P_{3/2}(F'=3) - 5D_{5/2}(F''=4)$ transition. As shown in Fig. 1(a), all the laser beams are linearly polarized, but two polarization configurations are considered. In the first (parallel) scheme, the polarization vectors of the probe, the first, and second coupling beams are in the $x$, $y$, and $y$ directions, respectively. In the second (perpendicular) scheme, the polarization vector of the second coupling beam is switched to the $x$ direction. In Fig. 1(a), the double-sided arrows and the dots denote the direction of the electric field of the laser beams. Thus, the polarizations of the coupling beams are parallel (perpendicular) for the first (second) configuration. When the polarization direction of the probe beam is chosen as the quantization axis, the energy-level diagrams for the parallel and perpendicular polarization schemes are those shown in Figs. 1(c) and 1(d), respectively. The solid red, solid green, and dotted green lines represent the probe, the first, and the second coupling beams, respectively. As can be seen in these figures, the transition schemes for the two configurations are completely different. In the parallel configuration, the two coupling beams are tuned to the same transition lines, whereas those beams are tuned to the different transition lines in the perpendicular configuration. The analytical calculation of the absorption coefficients for the basic units of these two schemes is discussed in the next section.

\begin{figure}[thb]
\centerline{\includegraphics[width=8cm]{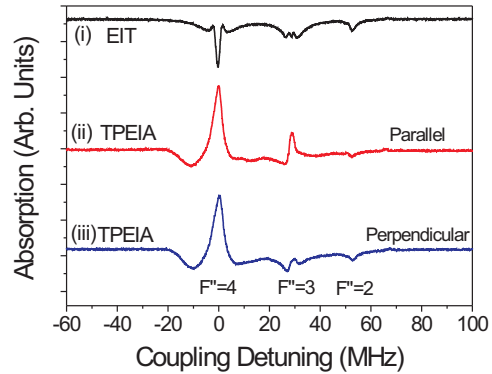}} \caption{(Color online) (i) An experimental EIT spectrum. Typical experimental TPEIA spectra for (ii) the parallel and (iii) the perpendicular polarization configurations.}
\end{figure}

Figure 2 shows typical experimental results for EIT and TPEIA. We refer to previous papers for the details of the experimental setup \cite{Jeong,YSLee}. The powers of the probe and of the first and second coupling beams were 8 $\mu$W, 20 mW, and 20 mW, respectively. The diameter of the laser beams was 2 mm. In Fig. 2, we present two TPEIA spectra for the parallel and perpendicular polarization configurations, and an EIT spectrum obtained in the absence of the second coupling beam. In Fig. 2, the frequency of the probe beam was tuned to the resonant transition $5S_{1/2}(F=2) - 5P_{3/2}(F'=3)$, whereas the frequencies of the coupling beams were scanned around the transition $5P_{3/2}(F'=3)-5D_{5/2}(F''=2,3,4)$. As discussed previously \cite{Jeong,YSLee}, the EIT signal transformed into the absorption signal when the second coupling beam was introduced. As can be seen in Fig. 2 for the two polarization configurations, the two TPEIA spectra for the $5S_{1/2}(F=2) - 5P_{3/2}(F'=3)-5D_{5/2}(F''=4)$ transition are very similar. As mentioned above, the interaction energy level diagrams for these two schemes are very different. However, we observed very similar TPEIA spectra for the two schemes, as can be seen in Fig. 2. This is the motivation of our work.

\section{Theory}
\label{theory}

\begin{figure*}[hbt]
\centerline{\includegraphics[width=14cm]{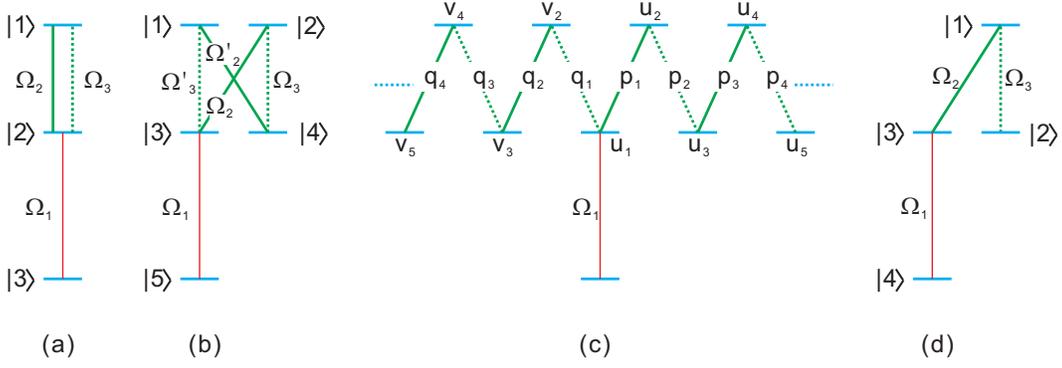}} \caption{(Color online) Simple energy-level diagram for (a) three-level and (b) five-level schemes. (c) The energy-level diagram with multiple alternative interactions for the upper transition line, equivalent to the three-level scheme. (d) Four-level scheme.}
\end{figure*}

To study the similarity and difference between the results of TPEIA for the two polarization configurations analytically, we used the two simplified energy-level diagrams shown in Fig. 3. The transitions (Rabi frequency) of the probe, and of the first and second coupling beams are denoted by solid red ($\Omega_1$), solid green ($\Omega_2$), and dotted green lines ($\Omega_3$), respectively. Figures 3(a) and 3(b) show the basic units of the diagrams corresponding to the parallel and perpendicular-polarization configurations, respectively. In Fig. 3(b), $\Omega'_2$ ($\Omega'_3$) denotes the Rabi frequency of the first (second) coupling beam for the other transition line. Figure 3(c) shows the energy-level diagram with multiple alternative interactions for the upper transition line. This diagram was proved to be equivalent to the schemes in Figs. 3(a) and (b) and will be discussed below in detail. Figure 4(d) shows the four-level scheme, the simplest primitive scheme for the two schemes in Figs. 3(a) and (b).

We calculated analytically the absorption coefficients of the probe beam in the weak-probe-intensity limit, for an arbitrary intensity of the coupling beams. We first considered the three-level scheme shown in Fig. 3(a). The decay rates of the states $\left|1\right>$, $\left|2\right>$, and $\left|3\right>$ are $\Gamma_1$, $\Gamma_2$, and $\Gamma_3 (= 0)$, and the decay rates of the optical coherences between the states $\left| \mu \right>$ and $\left| \nu \right>$ are $\gamma_{\mu \nu} = \frac{\Gamma_\mu +\Gamma_\nu}{2}$ with $\mu <\nu$. The wavelengths, wave vectors, and detunings of the probe (coupling) beam are $\lambda_1$, $k_1$, and $\delta_p$ ($\lambda_2$, $k_2$, and $\delta_c$), respectively. The effective detunings (experienced by an atom moving at velocity $v$) of the probe and of the first and second coupling beams are $\delta_1 = \delta_p -k_1 v$, $\delta_2 = \delta_c +k_2 v$, and $\delta_3 = \delta_c -k_2 v$, respectively.

We considered the solutions up to the first order in the probe Rabi frequency. In the energy level diagram shown in Fig. 3(a), the two coupling beams were tuned to the transition between $\left| 1\right>$ and $\left|2\right>$. Therefore, the frequency mixing produced a variety of oscillation frequencies for the optical coherences and populations. Because of the weak probe intensity, the optical coherence $\rho_{12}$ can be neglected, and thus we need consider only the optical coherences $\rho_{23}$ and $\rho_{13}$. In addition, all the stationary and oscillating terms in the populations vanish except for the stationary term of $\rho_{33}$, which is unity in the limit of weak probe intensity. If the probe beam is not quite weak, the coherent population oscillation (CPO) in the populations of the ground, intermediate, and excited states should be considered \cite{CPO1,CPO2}. In this case we would need a numerical calculation instead of the analytical calculation. We refer to previous publications for the method of calculation of the oscillation frequencies \cite{Noh07,Noh1,Noh2}. When $N$-photon interactions are taken into account, the optical $\rho_{23}$ and $\rho_{13}$ are explicitly given by
\begin{subequations}
\begin{eqnarray}
\rho_{23}&=&u_1 +u_{3}e^{i \delta_d t}+v_{3}e^{-i \delta_d t}+u_{5}e^{2 i \delta_d t}+v_{5}e^{-2 i \delta_d t} \nonumber \\
&& + \cdots +u_{N}e^{i \frac{N-1}{2} \delta_d t}+v_{N}e^{-i \frac{N-1}{2} \delta_d t}, \label{rho23}\\
\rho_{13}&=& u_{2}+v_{2}e^{-i \delta_d t}+u_{4}e^{i \delta_d t}+v_{4}e^{-2 i \delta_d t}+ \cdots \nonumber \\
&& +u_{N-1}e^{i \frac{N-3}{2} \delta_d t}+v_{N-1}e^{-i \frac{N-1}{2} \delta_d t},\label{rho13}
\end{eqnarray}
\end{subequations}
where $\delta_d \equiv \delta_3 -\delta_2=-2k_2 v$ and $N$ is assumed to be an odd integer. $u$ and $v$ in Eqs. (\ref{rho23}) and (\ref{rho13}) are Fourier components, which should be obtained by solving the density matrix equations, and the subscripts of $u$ and $v$ represent the number of photon interactions. Thus, the Fourier components of $\rho_{23}$ ($\rho_{13}$) result from an odd (even) number of photon interactions. From the density matrix equations, we obtained the differential equations for the components of $\rho_{23}$ and $\rho_{13}$ as follows. The lowest-order equation for ${\dot u}_{1}$ is given by
\begin{equation}\label{firstu}
{\dot u}_{1} = i \Delta_1 u_{1}  -\frac{i}{2} \Omega_1  -\frac{i}{2} \Omega_2 u_{2}-\frac{i}{2} \Omega_3 v_{2}.
\end{equation}
We can see that the equations for the $u$- and $v$-series can be constructed separately. The equations for the $u$-series are given by
\begin{eqnarray}
{\dot u}_{2} &=& i \Delta_2 u_{2}-  i p_1 u_{1}-i p_2 u_{3}, \nonumber  \\
{\dot u}_{3} &=& i \Delta_3 u_{3}-  i p_2 u_{2}-i p_3 u_{4}, \nonumber  \\
&&\cdots \nonumber \\
{\dot u}_{n} &=& i \Delta_n u_{n}-  i p_{n-1} u_{n-1}-i p_{n} u_{n+1}, \label{unpuls1}  \\
&&\cdots  , \nonumber
\end{eqnarray}
and those for the $v$-series by
\begin{eqnarray}
{\dot v}_{2} &=& i \Delta'_2 v_{2}-  i q_1 u_{1}-i q_2 v_{3}, \nonumber  \\
{\dot v}_{3} &=& i \Delta'_3 v_{3}-  i q_2 v_{2}-i q_3 v_{4}, \nonumber  \\
&&\cdots \nonumber \\
{\dot v}_{n} &=& i \Delta'_n v_{n}-  i q_{n-1} v_{n-1}-i q_{n} v_{n+1},  \label{vnpuls2} \\
&&\cdots  ,  \nonumber
\end{eqnarray}
where $v_1 = u_1$, and
\begin{eqnarray}
p_n (q_n) = \left\{ \begin{array}{ll}  \frac{\Omega_2}{2} \left(\frac{\Omega_3}{2}\right)  \, , &  {\rm for}\;\; n= 1,3,5, \cdots  \\   \\
\frac{\Omega_3}{2} \left(\frac{\Omega_2}{2}\right) \, , & {\rm for}\;\; n=2,4,6,\cdots .
  \end{array}
  \right. \label{pnqn1}
\end{eqnarray}
The effective detunings are given by
\begin{eqnarray}
\Delta_{2n} &=& \delta_1 +n \delta_2 -(n-1)\delta_3+i \gamma_{13}, \nonumber \\
\Delta_{2n+1} &=& \delta_1 +n \delta_2 -n \delta_3 +i \gamma_{23}, \nonumber \\
\Delta'_{2n} &=& \delta_1 +n \delta_3 -(n-1)\delta_2+i \gamma_{13}, \nonumber \\
\Delta'_{2n+1} &=& \delta_1 +n \delta_3 -n \delta_2 +i \gamma_{23}. \nonumber
\end{eqnarray}
Equations (\ref{firstu})--(\ref{vnpuls2}) can be described graphically as in Fig. 3(c). As we discuss below, the results for the schemes in Figs. 3(a) and (b) are equivalent to the energy-level diagram in Fig. 3(c). The right and left branches denote the equations for the $u$- and $v$-series, respectively. Since these equations are equivalent under exchange of $\Omega_2 (\delta_2)$ and $\Omega_3(\delta_3)$, we need only solve the solutions for either the $u$- or $v$-series. Here, we solve the former.

We assumed the following relation:
\begin{eqnarray}\label{unsimple}
u_n = \frac{p_{n-1}}{A_n}u_{n-1}.
\end{eqnarray}
If we insert Eq. (\ref{unsimple}) into Eq. (\ref{unpuls1}) by replacing $n$ with $n+1$, i.e., $u_{n+1}=\frac{p_{n}}{A_{n+1}}u_{n}$, we obtain
\begin{eqnarray}\label{unsimple1}
u_{n}=\frac{p_{n-1}}{\Delta_{n}-\frac{p_{n}^2}{A_{n+1}}}u_{n-1}.
\end{eqnarray}
Comparison of Eqs. (\ref{unsimple}) and (\ref{unsimple1}) yields the following recursion relation:
\begin{eqnarray}\label{An}
A_{n}=\Delta_{n}-\frac{p_{n}^2}{A_{n+1}}.
\end{eqnarray}
Because the maximum value of $n$ is $N$, $A_N=\Delta_N$ and $A_n=0$ when $n>N$.

Using Eqs. (\ref{pnqn1}), (\ref{unsimple}), and (\ref{An}), we can express $u_2$ in terms of $u_1$ as
\begin{equation}\label{finalu2}
u_2 =u_1
\cdot \frac{\Omega_2}{2} \left[ \Delta_2 -\frac{\Omega_3^2 /4}{\Delta_3 -\frac{\Omega_2^2 /4}{\Delta_4 -\frac{\Omega_3^2 /4}{\Delta_5 \cdots}}} \right]^{-1}  .
\end{equation}
Similarly for $v_2$,
\begin{equation}\label{finalv2}
v_2 =u_1
\cdot \frac{\Omega_3}{2} \left[ \Delta'_2 -\frac{\Omega_2^2 /4}{\Delta'_3 -\frac{\Omega_3^2 /4}{\Delta'_4 -\frac{\Omega_2^2 /4}{\Delta'_5 \cdots}}} \right]^{-1}  .
\end{equation}
Inserting Eqs. (\ref{finalu2}) and (\ref{finalv2}) into Eq. (\ref{firstu}), we obtain the final result for $u_1$:
\begin{eqnarray}\label{finalresult}
&& u_{\|}=\frac{\Omega_1}{2}  \nonumber \\ && \times \left[ \Delta_1 - \frac{\Omega_2^2/4}{\Delta_2 -\frac{\Omega_3^2 /4}{\Delta_3 -\frac{\Omega_2^2 /4}{ \Delta_4 -\frac{\Omega_3^2 /4}{\Delta_5 \cdots}}}} -\frac{\Omega_3^2 /4}{ \Delta'_2 -\frac{\Omega_2^2 /4}{ \Delta'_3 -\frac{\Omega_3^2 /4}{ \Delta'_4 -\frac{\Omega_2^2 /4}{\Delta'_5 \cdots}}}} \right]^{-1},
\end{eqnarray}
where $u_1$ for the parallel polarization configuration is denoted $u_{\|}$. We notice that the terms associated with $\Delta_3$ and $\Delta'_3$ are responsible for TPEIA and result from the three-photon coherences $u_3$ and $v_3$ in Eq. (\ref{rho23}), respectively. As shown in the next section, when thermal averaging is considered, the contribution of $u_3$ is much larger than that of $v_3$ owing to the Doppler-free characteristics. It should be noted also that the term $u_3$ in Eq. (\ref{rho23}), responsible for the four-wave mixing signal, is given by
\begin{eqnarray}
u_3 = \frac{\Omega_2 \Omega_3}{4} \times \frac{u_{\|}}{\Delta_2 \left( \Delta_3 -\frac{\Omega_2^2 /4}{\Delta_4 -\frac{\Omega_3^2 /4}{\Delta_5 \cdots} } \right) -\frac{\Omega_3^2}{4}}.
\end{eqnarray}

We performed the calculation for the five-level model in Fig. 3(b), the basic unit for the energy-level diagram for the perpendicular polarization configuration. As in the three-level scheme, we considered the solutions up to the first order in the probe Rabi frequency. Thus, we need only consider the elements $\rho_{35}$, $\rho_{45}$, $\rho_{15}$, $\rho_{25}$, and $\rho_{55} (\simeq 1)$ while neglecting all the other elements. From the calculation of the oscillation frequencies, these density-matrix elements can be expanded as follows:
\begin{subequations}
\begin{eqnarray}
\rho_{35}&=&u_1 +u_{5}e^{2 i \delta_d t}+v_{5}e^{-2 i \delta_d t}\nonumber \\
&& +u_{9}e^{4 i \delta_d t}+v_{9}e^{-4 i \delta_d t} + \cdots , \label{rho35}\\
\rho_{45}&=& u_{3}e^{ i \delta_d t}+v_{3}e^{-i \delta_d t}  \nonumber \\
&& +u_{7}e^{3 i \delta_d t}+v_{7}e^{-3 i \delta_d t} + \cdots , \label{rho45}\\
\rho_{25}&=& u_{2}+v_{4}e^{-2 i \delta_d t} \nonumber \\
&& +u_{6}e^{2 i \delta_d t}+v_{8}e^{-4 i \delta_d t} + \cdots,\label{rho25}\\
\rho_{15}&=& v_{2}e^{-i \delta_d t}+u_{4}e^{i \delta_d t}  \nonumber \\
&& +v_{6}e^{-3 i \delta_d t}+u_{8}e^{3 i \delta_d t} + \cdots.\label{rho15}
\end{eqnarray}
\end{subequations}
Comparing Eqs. (\ref{rho35})--(\ref{rho15}) with Eqs. (\ref{rho23}) and (\ref{rho13}), it is easy to recognize that the terms in $\rho_{35}$ and $\rho_{45}$ are the same as those in $\rho_{23}$ in Eq. (\ref{rho23}). In addition, the terms in $\rho_{25}$ and $\rho_{15}$ are the same as those in $\rho_{13}$ in Eq. (\ref{rho13}). Thus, the structures of the density matrix elements of the three- and five-level schemes are identical.

When we solve the density-matrix equations using the expanded elements in Eqs. (\ref{rho35})--(\ref{rho15}), we obtain equations identical to those shown above in the discussion of the three-level scheme. The only difference is in the definitions of $p_n$ and $q_n$ because of the different Rabi frequencies that are due to the different transition strengths. In this case, $p_n$ and $q_n$ are defined as
\begin{eqnarray}
p_n (q_n) = \left\{ \begin{array}{ll}  \frac{\Omega_2}{2} \left(\frac{\Omega'_3}{2}\right)  \, , &  {\rm for}\;\; n= 1,5, 9, \cdots , \\ \\
\frac{\Omega_3}{2} \left(\frac{\Omega'_2}{2}\right) \, , & {\rm for}\;\; n=2,6,10,\cdots, \\ \\
\frac{\Omega'_2}{2} \left(\frac{\Omega_3}{2}\right)  \, , &  {\rm for}\;\; n= 3,7,11, \cdots  ,\\  \\
\frac{\Omega'_3}{2} \left(\frac{\Omega_2}{2}\right) \, , & {\rm for}\;\; n=4,8,12,\cdots .
  \end{array}
  \right. \nonumber
\end{eqnarray}
Then, the final value of $u_1$ for the perpendicular configuration is given by
\begin{eqnarray}\label{finalresultv}
&& u_{\bot} =\frac{\Omega_1}{2} \nonumber \\ && \times \left[ \Delta_1 - \frac{\Omega_2^2/4}{\Delta_2 -\frac{\Omega_3^2 /4}{\Delta_3 -\frac{{\Omega'}_2^2 /4}{\Delta_4 -\frac{{\Omega'}_3^2 /4}{\Delta_5 \cdots}}}} -\frac{{\Omega'}_3^2 /4}{\Delta'_2 -\frac{{\Omega'}_2^2 /4}{\Delta'_3 -\frac{\Omega_3^2 /4}{\Delta'_4 -\frac{\Omega_2^2 /4}{\Delta'_5 \cdots}}}} \right]^{-1} ,
\end{eqnarray}
where $u_1$ for the perpendicular-polarization configuration is denoted $u_{\bot}$. Equation (\ref{finalresultv}) is equivalent to Eq. (\ref{finalresult}) except for the different values of the Rabi frequencies. It should be noted that the three-photon coherence $u_3$ in Eq. (\ref{rho45}) is responsible for TPEIA.

As discussed in the next section, the terms of the $v$-series in Eqs. (\ref{finalresult}) and (\ref{finalresultv}) are not Doppler-free, so that the most primitive scheme for the parallel and perpendicular schemes is the simple four-level scheme shown in Fig. 3(d). The result for this scheme is simply given by
\begin{eqnarray}\label{finalresultN}
u_{\rm 4L} =\frac{\Omega_1}{2} \left[ \Delta_1 - \frac{\Omega_2^2/4}{\Delta_2 -\frac{\Omega_3^2 }{4\Delta_3 }}  \right]^{-1},
\end{eqnarray}
where $u_{\rm 4L}$ is $u_1$ for the four-level scheme.

Finally, the absorption coefficient for each polarization configuration averaged over the Doppler broadened velocity distribution is given by
\begin{eqnarray}\label{alphapara}
\alpha_{j} = -\frac{1}{\sqrt{\pi} v_{\rm mp}} \frac{3 \lambda_1^2}{2 \pi} \frac{N_{\rm at} \Gamma_2}{\Omega_1} \int_{-\infty}^{\infty} {\rm d}v \; e^{-\left(v/v_{\rm mp}\right)^2} {\rm Im} \left( u_j \right)   ,
\end{eqnarray}
where $j \in \{ \|, \bot$, ${\rm 4L}\}$, $v_{\rm mp}$ is the most probable speed, and $N_{\rm at}$ is the atomic density in the vapor cell.

\section{Results and Discussion}
\label{results}

\begin{figure}[thb]
\centerline{\includegraphics[width=8cm]{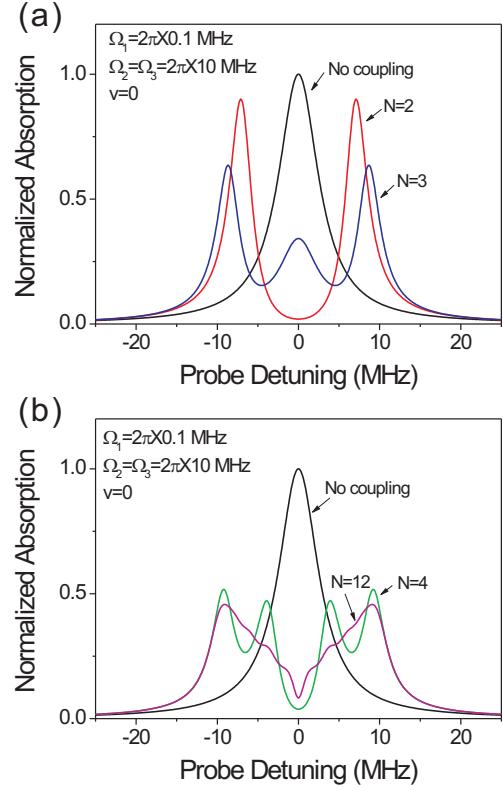}} \caption{(Color online) Calculated normalized absorptions for a stationary atom at (a) $N=1,2,$ and 3 and (b) $N=1,4,$ and 12.}
\end{figure}

The calculated normalized absorptions for a stationary atom in the parallel-polarization configuration with various values of $N$ are presented in Fig. 4. When $N=1$, the absorption is the background value in the absence of the two coupling beams. We obtained an EIT spectrum for $N=2$ where the second coupling beam is absent, and a TPEIA spectrum for $N=3$ in the presence of two coexisting coupling beams. In Fig. 4, as $N$ increases, the transmittance and absorption vary in an alternating fashion \cite{Pandey13}. As will be seen later, there is a big difference between the results for $N=2$ and $N=3$, but no significant differences for the higher values of $N$.

\begin{figure}[thb]
\centerline{\includegraphics[width=8cm]{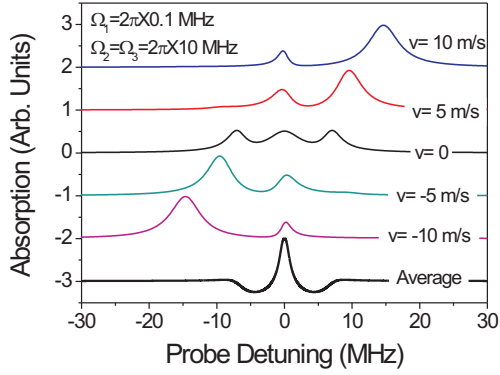}} \caption{(Color online) Calculated absorptions for an atom with the four-level scheme, moving at velocity $v=0$, $\pm 5$, or $\pm 10$ m/s. The bottom curve refers to the thermal-averaged absorption spectrum.}
\end{figure}

In order to study the effect of thermal averaging, we present the absorption coefficient for the four-level scheme depicted in Fig. 3(d). We used this scheme because this is the most primitive in the exact energy-level diagrams of the parallel and perpendicular configurations. In addition, it can simplify the occurrence of absorption resonances. The four-level scheme corresponds to $N=3$ where only the $u$-branch exists. This choice will be validated soon. Figure 5 plots the normalized absorptions for the velocities $v=0$, $\pm 5$, and $\pm 10$ m/s, and the thermal-averaged absorption. The resonances occur at $\delta_p=0$ and $\pm (1/2)\sqrt{\Omega_2^2 +\Omega_3^2}$ when $v=0$, from a dressed-state analysis. As the magnitude of the velocity increases, the two side resonances shift far away from the origin, but the central resonance remains almost stationary at the origin. Therefore, after averaging over the velocity distribution, the central peak survives, while the two side resonances cancel out. Although the analysis was performed for the four-level scheme, this analysis is valid for the other complicated diagrams in Fig. 3.

\begin{figure}[thb]
\centerline{\includegraphics[width=8cm]{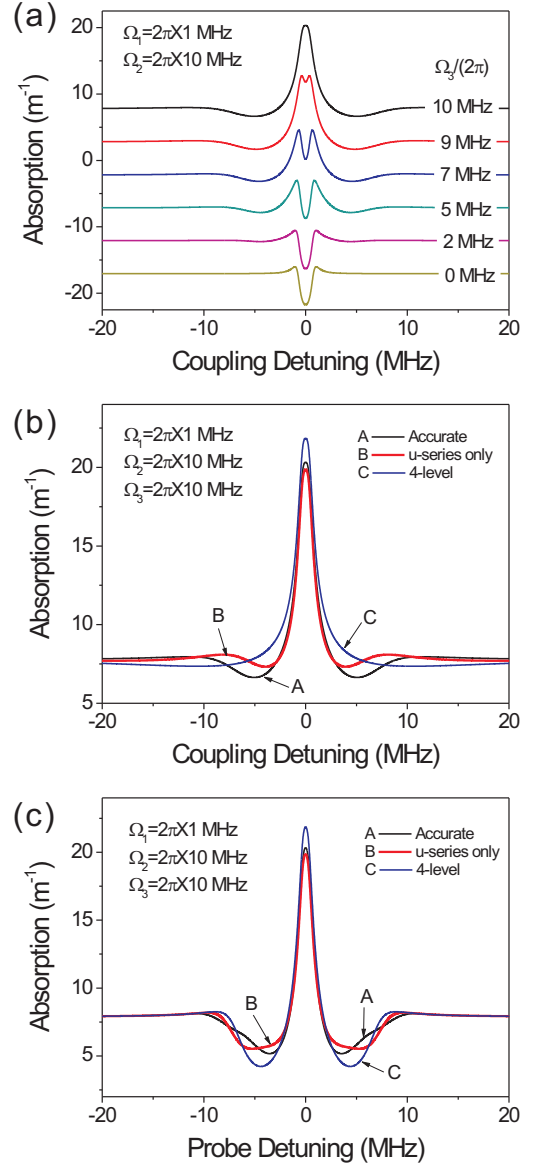}} \caption{(Color online) (a) TPEIA spectra for various $\Omega_3$ values. TPEIA spectra obtained via three different calculating methods with scanning of (b) coupling detuning and (c) probe detuning.}
\end{figure}

Figure 6 shows the series of averaged absorption spectra for the analytical results in Eq. (\ref{alphapara}). Figure 6(a) shows the absorption spectra for several values of the Rabi frequency of the second coupling beam, where $\Omega_1 = 2 \pi \times 1$ MHz and $\Omega_2 = 2 \pi \times 10$ MHz. As can be seen in Fig. 6(a), the EIT spectrum at $\Omega_3 =0$ transforms to the TPEIA spectrum as $\Omega_3$ increases. Figure 6(b) shows three typical spectra for the parallel-polarization configuration where $\Omega_1 = 2 \pi \times 1$ MHz, $\Omega_2 =\Omega_3 = 2 \pi \times 10$ MHz, and the probe detuning was set to zero. The black curve (A) denotes the exact result of Eqs. (\ref{finalresult}) and (\ref{alphapara}). The red curve (B) denotes the spectrum where the term of the $v$-series was neglected. Clearly, the two spectra are very similar. This means that the contribution of the $v$-series is much smaller than that of the $u$-series. The $u$- and $v$-series are represented graphically as the right and left branches in Fig. 3(c). The reason for the weak contribution of the $v$ series is that it is not Doppler-free. In Eq. (\ref{finalresult}), the two-photon resonance terms for the $u$- and $v$-series are $\Delta_2 = \delta_1+\delta_2 +i \gamma_{13}$ and $\Delta'_2 =\delta_1+\delta_3 +i \gamma_{13}$, respectively. Because $\delta_1+\delta_2 = \delta_p+\delta_c +\left(k_2 - k_1 \right)v$ and $\delta_1+\delta_3 = \delta_p+\delta_c -\left(k_2 + k_1 \right)v$, the term $\Delta_2$ contributes much more than $\Delta'_2$ after averaging over the velocity distribution. The blue curve (C) in Fig. 6(b) shows the result for the four-level scheme depicted in Fig. 3(d). This is very similar to the exact calculated spectra. As $\Omega_2$ and $\Omega_3$ decrease, the discrepancy between the four-level and exact results decrease. Figure 6(c) shows the same spectra as in Fig. 6(b), but with the probe detuning being scanned while the coupling detuning is set to zero. The overall spectra in Figs. 6(b) and (c) show very similar trends.

Finally we consider the conditions for detunings of the two coupling beams to create TPEIA signal, which were set to zero in the paper. Now we assume that the coupling detunings are arbitrary, and thus $\delta_2 = \delta_{c1}+k_2 v$ and $\delta_3 = \delta_{c2}-k_2 v$ where $\delta_{c1}$ and $\delta_{c2}$ are the detunings of the first and second coupling beams. Equation (\ref{finalresultN}) can be written as
\begin{equation}\label{condition}
u_{\rm 4L}=\frac{\Omega_1}{2} \times \frac{4 \Delta_2 \Delta_3 -\Omega_3^2}{4 \Delta_1 \Delta_2 \Delta_3 - \left( \Delta_3 \Omega_2^2 +\Delta_1 \Omega_3^2 \right)}.
\end{equation}
In Eq. (\ref{condition}), we have two conditions for the two-photon Doppler-free characteristics such as $\Delta_2 =0$ and $\Delta_3 \Omega_2^2 +\Delta_1 \Omega_3^2=0$. These conditions result in the requirements $\delta=-\delta_{c1}$ and $\delta_{c1}=-\delta_{c2}$ when $k_1 \simeq k_2$ and $\Omega_2 \simeq \Omega_3$. Therefore, the frequencies of the first and second coupling beams should be located at the symmetric positions relative to the upper resonant transition line. In our case, because $\delta_{c1}=\delta_{c2} =0$, we see the TPEIA signal at the zero probe detuning as shown in Fig. 5.

\section{Conclusion}
\label{conclusion}

We have presented a theoretical study of three-photon electromagnetically induced absorption in ladder-type atomic systems. The primitive energy-level diagrams for the parallel and perpendicular coupling-beam polarization configurations were the three- and five-level schemes, respectively. We derived analytically the absorption coefficients for these two schemes and found them to be equivalent to a ladder system with an upper transition of multiple alternative interacting schemes. The absorption coefficient consisted of contributions from the two opposite branches, one of which was Doppler-free. It was also found that the simplest scheme was the simple four-level scheme. 

In contrast to normal EIA in degenerate two-level systems, where the transfer of coherence or population plays a major role \cite{Goren03,Goren04}, TPEIA involves constructive interference \cite{Mulchan00,Gong11} and three-photon coherence is the main cause of TPEIA in both parallel and perpendicular schemes. Although the basic mechanism of TPEIA emerged from the analytical calculation of the absorption spectra, TPEIA for real atoms displayed interesting behavior. For example, the polarization dependence was different depending on whether or not the transition was closed. An accurate calculation and analysis for TPEIA for real atoms are currently in progress.
 
\acknowledgements

This research was supported by Basic Science Research Program through the National Research Foundation of Korea(NRF) funded by the Ministry of Science, ICT and future Planning(2014R1A2A2A01006654 and 2015R1A2A1A05001819).

%\centerline{}

%\newpage

%\centerline{\bf APPENDIX}

%\setcounter{equation}{0}
%\renewcommand{\theequation}{A{\arabic{equation}}}

%\newpage

%\vspace{2cm}

%\centerline{\bf FIGURES}

%\vskip.5in

%\begin{itemize}

%\item FIG. 1: (Color online) (a) Simplified schematic diagram for the TPEIA experiment. (b) Energy-level diagram for the $5S_{1/2}-5P_{3/2}-5D_{5/2}$ transition in $^{87}$Rb atoms. The two-coupling beams are linearly polarized in (c) parallel or (d) perpendicular direction.

%\centerline{}

%\item FIG. 2: (Color online) (i) An experimental EIT spectrum. Typical experimental TPEIA spectra for (ii) the parallel and (iii) the perpendicular polarization configurations.

%\centerline{}

%\item FIG. 3: (Color online) Simple energy-level diagram for (a) three-level and (b) five-level schemes. (c) The energy-level diagram with multiple alternative interactions for the upper transition line, equivalent to the three-level scheme. (d) Four-level scheme.

%\centerline{}

%\item FIG. 4: (Color online) Calculated normalized absorptions for a stationary atom at (a) $N=1,2,$ and 3 and (b) $N=1,4,$ and 12.

%\centerline{}

%\item FIG. 5: (Color online) Calculated absorptions for an atom with the four-level scheme, moving at velocity $v=0$, $\pm 5$, or $\pm 10$ m/s. The bottom curve refers to the thermal-averaged absorption spectrum.

%\centerline{}

%\item FIG. 6: (Color online) (a) TPEIA spectra for various $\Omega_3$ values. TPEIA spectra obtained via three different calculating methods with scanning of (b) coupling detuning and (c) probe detuning.

%\end{itemize}

\end{document}